\documentclass[sigconf]{acmart}
\usepackage{booktabs} 
\usepackage{multirow}
\usepackage{amsmath}
\usepackage{mathtools}

\usepackage{amssymb}
\usepackage{graphics}
\usepackage{subfigure}
\usepackage{lipsum}

\usepackage{bm}
\usepackage{bbding}
\usepackage{pifont}
\usepackage{wasysym}
\usepackage{balance}

\usepackage{array}
\newcolumntype{H}{>{\setbox0=\hbox\bgroup}c<{\egroup}@{}}

\newcommand{\commentout}[1]{}

\newcommand{\figref}[1]{Fig.~\ref{#1}}

\newcommand{\tblref}[1]{Tbl.~\ref{#1}}

\newtheorem{definition}{Definition}
\newtheorem{theorem}[definition]{Theorem}
\newtheorem{postulate}[definition]{Postulate}
\newtheorem{proposition}[definition]{Proposition}
\newtheorem{lemma}[definition]{Lemma}
\newtheorem{corollary}[definition]{Corollary}
\newtheorem{conjecture}[definition]{Conjecture}

\DeclareSymbolFont{libertineg}{\encodingdefault}{\familydefault}{m}{it}
\SetSymbolFont{libertineg}{bold}{\encodingdefault}{\familydefault}{b}{it}
\DeclareMathSymbol{g}{\mathalpha}{libertineg}{`g}

\copyrightyear{2023}
\acmYear{2023}
\setcopyright{acmlicensed}
\acmConference[KDD '23] {Proceedings of the 29th ACM SIGKDD Conference on Knowledge Discovery and Data Mining}{August 6--10, 2023}{Long Beach, CA, USA.}
\acmBooktitle{Proceedings of the 29th ACM SIGKDD Conference on Knowledge Discovery and Data Mining (KDD '23), August 6--10, 2023, Long Beach, CA, USA}
\acmPrice{15.00}
\acmISBN{979-8-4007-0103-0/23/08}
\acmDOI{10.1145/3580305.3599914}

\begin{document}


\title{Towards Disentangling Relevance and Bias in Unbiased Learning to Rank}

\author{%
 Yunan Zhang, Le Yan, Zhen Qin, Honglei Zhuang, Jiaming Shen, Xuanhui Wang, Michael Bendersky, Marc Najork
}
\affiliation{ 
 \institution{University of Illinois Urbana-Champaign, Google Research}
}
\email{yunanz2@illinois.edu, {lyyanle,zhenqin,hlz,jmshen,xuanhui,bemike,najork}@google.com}

\begin{abstract}

Unbiased learning to rank (ULTR) studies the problem of mitigating various biases from implicit user feedback data such as clicks, and has been receiving considerable attention recently. A popular ULTR approach for real-world applications uses a two-tower architecture, where click modeling is factorized into a relevance tower with regular input features, and a bias tower with bias-relevant inputs such as the position of a document. A successful factorization will allow the relevance tower to be exempt from biases. In this work, we identify a critical issue that existing ULTR methods ignored - the bias tower can be confounded with the relevance tower via the underlying true relevance. In particular, the positions were determined by the logging policy, i.e., the previous production model, which would possess relevance information. We give both theoretical analysis and empirical results to show the negative effects on relevance tower due to such a correlation. We then propose two methods to mitigate the negative confounding effects by better disentangling relevance and bias. Offline empirical results on both controlled public datasets and a large-scale industry dataset show the effectiveness of the proposed approaches. 
We conduct a live experiment on a popular web store for four weeks, and find a significant improvement in user clicks over the baseline, which ignores the negative confounding effect.

\end{abstract}


 \maketitle

\fancyhead{}

\section{Introduction}


Learning to rank (LTR) is critical for many real-world applications such as search and recommendations~\citep{8186875, dasalc}. In practice, ranking models are often trained using implicit feedback, e.g. user clicks, collected from serving logs. Though easily available in a large scale, implicit feedback has various kinds of biases, such as position bias. To address this issue, Unbiased Learning To Rank (ULTR) has gained much attention as it aims to mitigate these biases ~\cite{ai2021unbiased}.

Unbiased Learning To Rank (ULTR) methods can broadly be classified into two categories: counterfactual and click modeling approaches. Counterfactual approaches are mainly based on the Inverse Propensity Scoring (IPS) method~\cite{ai2018unbiased,10.1145/3488560.3498375, wang2016selectionbias, joachims2017unbiased}. These approaches require knowing the bias or observation propensities in advance, but they are not always necessarily available. The second category covers click modeling approaches that achieve relevance estimation through factorized models trained to predict clicks. This type of approaches do not require observation propensities in advance. In this paper, we focus on the second category, and in particular, the two-tower additive models, which are popular in industrial ranking systems~\cite{youtube, guo2019pal, haldar2020improving, chu2021general} due to simplicity and effectiveness. 

In the two-tower models, one tower takes regular input features to model unbiased relevance predictions, while the other tower takes bias-related features, such as position and platform (e.g., mobile vs desktop) to estimate users' non-uniform observation probability over the results. We thus use bias and observation interchangeably in this paper. During offline training, the outputs of these two towers are added together to explain logged user feedback. However, during online serving, only the unbiased prediction tower is effectively used due to the fact that no logged position is available during serving. Two-tower additive models are easy to implement and interpret: they follow the Position Based Model (PBM) click model~\cite{richardson2007predicting,chuklin2015click} to model user behaviors, which assumes that bias learning is independent of true relevance. Ideally, the factorization will automatically let the two-tower models learn relevance and biases into their respective towers. 

\begin{figure}[t]
\centering
\includegraphics[scale=0.6]{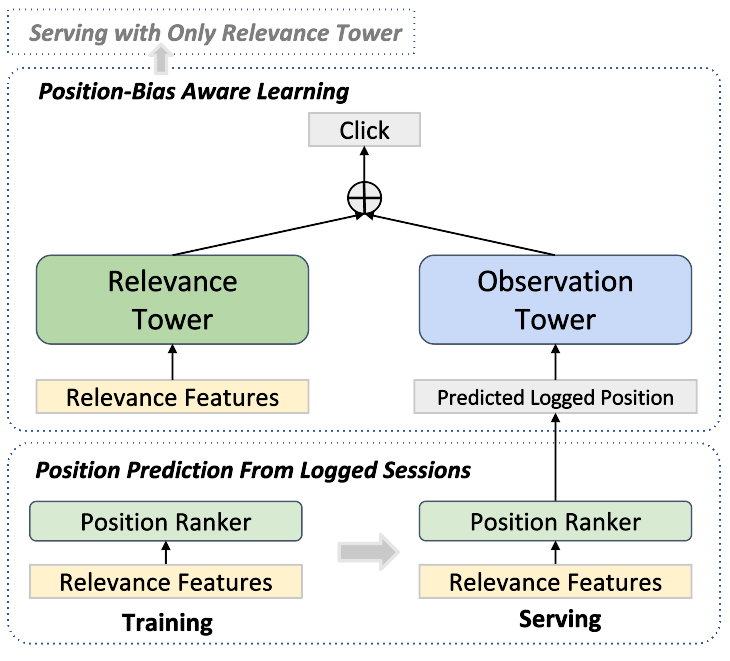}
\caption{An illustration of the two tower additive model ULTR framework. The input of the observation tower is generated from a ranker trained to learn relevance.}
\label{fig:twotower}
\end{figure}

However, we argue that such wishful thinking is unrealistic in practice. The key issue ignored in the literature is that, the relevance tower and observation tower are confounded by true relevance in terms of causal inference~\cite{pearl2009causality}. As shown in Figure \ref{fig:twotower}, the logged display position of an item was actually determined by the logging policy, i.e., the previous deployed ranker, which is likely correlated with true relevance. Thus, bias features and relevance features are both correlated with true relevance, which violates the independence assumption of observation and relevance signals. Overlooking such correlations between relevance features and observation features can be detrimental to the relevance tower to be served. Consider the extreme case where the logged positions were from a perfect relevance model (please note this is only an example for intuition), then the observation tower can potentially absorb \textit{all} relevance information during two-tower model learning and explain the clicks perfectly. The resulting  relevance tower, on the other hand, can be completely random and thus useless.

In this paper, we study the confounding effect both theoretically and empirically. We perform controlled semisynthetic experiments with different relevance-entangling levels by injecting different percentages of ground-truth information into the logging policy, from relevance-agnostic to total relevance-deterministic position data. The findings align with our intuition - the more correlated the bias and relevance are, the less effective the relevance tower is. We further propose two methods to mitigate the negative confounding effect to improve the relevance tower performance, including (1) an adversarial training method to disentangle the relevance information from the observation tower, (2) a dropout algorithm that learns to zero-out relevance-related neurons in the observation tower. We show that both proposed methods can disentangle the relevance and bias by mitigating the confounding issue, with different strengths and weaknesses. 
Moreover, we have conducted a live experiment of method (1) on a popular web store for four weeks and live experiment results also show significant improvements of the proposed methods in terms of user clicks.

In summary, the contributions of our work are three-fold:
\begin{itemize}
    \item We identify and analyze the confounding effect between relevance and bias in ULTR, which may bring detrimental effects to real-world ranking systems.  
    \item We propose two methods to mitigate this issue to disentangle relevance and bias and show promising results in both controlled semi-synthetic and real-world industry datasets.
    \item We provide a theoretical analysis on why the confounding factors can negatively affect relevance model learning, and on why our methods work.
    \item We conduct a live experiment on a popular web store for four weeks, and reveal significant improvements of proposed methods in terms of user clicks.
\end{itemize}

\section{The Problem}
In this section, we describe the general formulation and key assumptions of the two-tower ULTR model, and identify the confounding effect that can negatively affect the learning process. 

\begin{figure}[t]
\includegraphics[width=8.5cm]{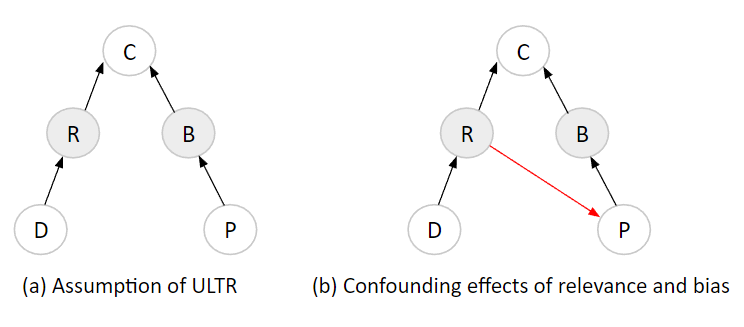}
\caption{The causal graphs of click modeling. \textit{C}-click, \textit{R}-true relevance, \textit{B}-bias, \textit{P}-position, \textit{D}-regular item features. Grey circles indicate unobserved latent variables. White circles indicate observed variables. }
\label{fig:causal}
\end{figure}

\subsection{General Assumption}
As shown in Figure~\ref{fig:causal}, in the two-tower model, a relevance-observation factorization is assumed for the click behavior. The click probability of an item $i$ can be factorized into two factors: the relevance of the item, and the observation probability of the item. The relevance of the item depends only on the item features and its interaction with the query $q$,  $\vec{x}_i$. The observation probability is only related to the display properties, e.g., position $p_i$. We can then formulate the click probability of item i as,
\begin{equation}
    P(C_i=1|\vec{x}_i,p_i) = P_{rel}(\vec{x}_i)\times P_{obs}(p_i).
\end{equation}
The two-tower models thus factorize the feature dependent relevance tower and the position dependent observation tower to model the clicks as,
\begin{equation}
\hat{c}_i = \psi(f(\vec{x}_i),g(p_i)),
\end{equation}
where the relevance tower $f$ and the observation tower $g$ encode $\vec{x}_i$ and $p_i$ into relevance score and observation score, respectively. $\psi$ is a transformation that fuses relevance scores and observation scores into click scores. Such formulation is expected to be bias-aware and to learn a position-agnostic relevance representation with the relevance encoder $f$.

\subsection{The Confounding Problem}
The position-agnostic relevance learning is based on the assumption that relevance and observation are independent, which allows us to factorize them. However, $p_i$ is the logged position of an item during historical online model deployment, where we had a ranker to rank items based on their predicted relevance $r$: 
\begin{equation}
    p_i = ||j| r(\vec{x}_j) > r(\vec{x}_i), \textrm{for} \: j\:  \textrm{in} \: [1, n]|| + 1
    \label{eqn:position-at-i}
\end{equation}
where $r(\vec{x}_i)$ is the relevance score for the item $\vec{x}_i$. The position $p_i$ is obtained by computing all the $n$ relevance scores associated with a query and sorting results by their scores in the descending order.

We can see from Equation \ref{eqn:position-at-i} that the position $p_i$ is determined by the estimated relevance in the logging policy. To what level the position reflects the true relevance depends on how the rank $r$ is correlated with the true relevance. As real-world industrial systems usually serve sensible rankers, the position could be a strong indication of relevance. In such case, the two-tower model learning process might find an obvious \textit{shortcut}~\cite{shortcut} to use position to explain relevance, since it can be easier to learn than the heavy (query, doc) features. Such learning behavior can be detrimental to the relevance tower learning. \par

A simple way to mitigate confounding bias is to perform online randomization experiments and gather data with logged position that is not confounded with relevance. However, it is not practical as randomization experiments hurt user experience significantly and real-world ranking systems usually need a lot of data to train. We focus on logged biased datasets in our methods below.

\section{Methods}
In this section, we introduce two methods to disentangle relevance and observation learning. The high-level idea is to control or unlearn the relevance information in the observation tower. We describe the model details below.

\subsection{DNN Architecture}
We describe the backbone two-tower DNN architecture shared by both methods.
\paragraph{Input Representation}
Each tower of the two-tower model takes one type of input: the relevance tower processes relevance-related features, while the observation tower handles observation-related features. The relevance input can be represented as a concatenation of various types of feature vectors depending on the data. We refer to the concatenated representation for item $i$ as $\vec{x}_i$. The observation features (an integer position index in the simplest case) are mapped to an embedding vector. If the observation tower input contains multiple features, we concatenate them in a similar way to relevance features. We still denote the observation input features as $p_i$ by slightly abusing the notation. 
\paragraph{Relevance Tower}
The relevance tower takes one item at a time. In this paper, we instantiate the relevance tower as a feed-forward neural network (FFN) $f$ whose parameters are denoted as $\theta$. Each layer of the network is fully connected with the ReLU activation and batch normalization. 
\begin{equation}
    f_\theta(\vec{x}_i) = FFN(\vec{x}_i; \theta)
\end{equation}
\paragraph{Observation Tower}
The observation tower takes a similar design to the relevance tower but smaller in size. In this paper, we instantiate the observation tower as a feed-forward neural network $g$ whose parameters are denoted as $\varphi$. Each layer of the network is fully connected with the ReLU activation and batch normalization. 
\begin{equation}
    g_\varphi(p_i) = FFN(p_i; \varphi)
\end{equation}
Here, $p_i$ is the logged ranking result of item $i$ by the serving ranker that was deployed in the production system during data collection.
\paragraph{Training}
The model is trained with the sigmoid cross-entropy loss supervised by user clicks. We denote the predicted click probability as $\hat{c}_i$,
\begin{equation}
\hat{c}_i = \psi(f_\theta(\vec{x}_i), g_\varphi(p_i)),
\end{equation}
where the interaction function $\psi$ is specific to each method, as shown below. 
With ground-truth click as ${c_i}$, we can optimize the following cross-entropy loss:
\begin{equation}
\label{eq:logloss}
\mathcal{L}_{click} =  -\sum_{i=1}^{n}[c_i\log(\hat{c}_i) + (1-c_i)\log(1-\hat{c}_i)].    
\end{equation}

\subsection{Gradient Reversal}
\label{sec:gradrev}
 In this method, the main click prediction is obtained as the additive model,
\begin{equation}
\label{eq:additive}
\psi(f, g) = {\rm sigmoid}(f + g).
\end{equation} 
The main issue we want to fix is the confounding relevance learned in the observation tower from the observation features. Inspired by research in adversarial learning and domain adaptation~\cite{ganin2015unsupervised}, we design a gradient reversal approach. The idea of gradient reversal in domain adaptation is to unlearn domain-specific information from the neural network by back-propagating the negative gradient when the model is trained with domain-specific labels.\par
\begin{figure}[h!]
\includegraphics[scale=0.25]{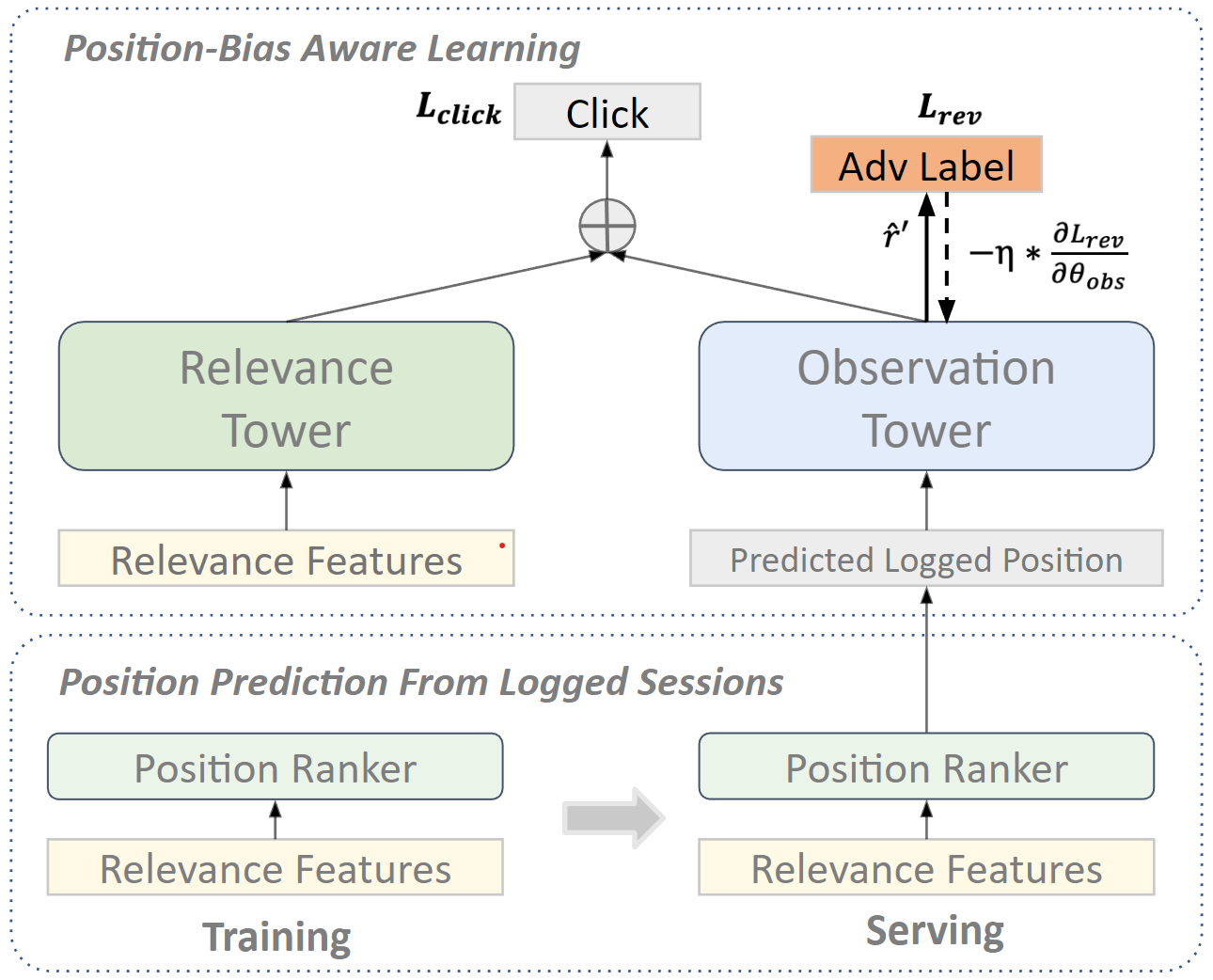}
\vspace{-0.8em}
\caption{An illustration of the gradient reversal method. The adv label provides supervision for the observation tower's gradient reversal task. The gradient of this task would be multiplied by a negative scaling factor $\eta$ in backpropagation.}
\label{fig:gradrev}
\end{figure}
In our case, we want to unlearn the relevance information captured by the observation tower, thus pushing more relevance information to the relevance tower during learning. So
we add an extra relevance prediction task to the observation tower. On top of the sharing hidden layers of the observation tower, we feed the observation tower output through a gradient reversal (GradRev) layer.  In the forward pass, the GradRev layer conducts an identity mapping of the input. In backpropagation, the layer reverses the gradient from the following layer, scales it by a hyperparameter $\eta$, and back-propagates it to the observation tower. The GradRev result then pass through a dense layer to predict its relevance output $g^{rev}$. We then supervise this output with an adversarial label, $y^{rev}$, that contains relevance-related signals, e.g., relevance tower predictions or clicks.
\begin{equation}
\label{eq:gradrev}
  \mathcal{L}_{rev} = \sum_{i=1}^{n}(y^{rev}_i - {\rm GradRev}(g^{rev}(p_i), \eta))^2.
\end{equation}
The total loss to be optimized is then the sum of $\mathcal{L}_{click}$ and $\mathcal{L}_{rev}$.
\par 
Figure \ref{fig:gradrev} gives an illustration of this method. We tried three different adversarial labels to prove the method's generality across scenarios. (1) Ground truth relevance label: this setting is mainly used for the ablation study, as true relevance is usually not available in user logs. (2) Click: we directly use user click as the adversarial label. (3) Relevance tower prediction:we use the predicted relevance score from the relevance tower. We will compare these choices in detail in Section \ref{sec:ablation-study}.

\subsection{Observation Dropout}
\label{sec:dropout}
\begin{figure}[h!]
\includegraphics[scale=0.3]{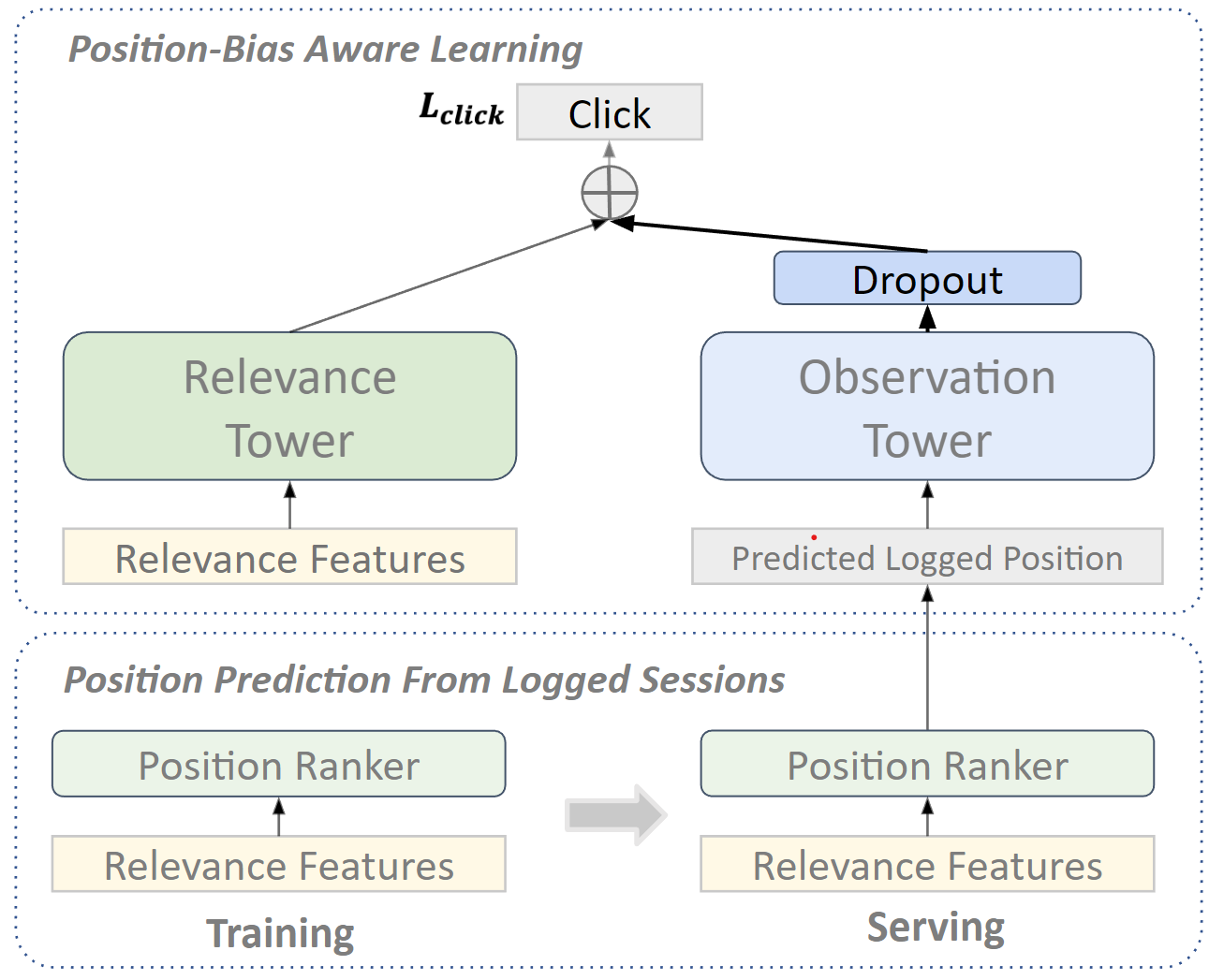}
\vspace{-0.8em}
\caption{Graphic illustration of the observation dropout method. A dropout layer is added to the observation score.}
\label{fig:dropout}
\end{figure}
Another perspective of the confounding issue is a classic shortcut learning problem common in neural networks~\cite{shortcut,bias,bugs}. The taxonomy, shortcut, refers to the scenario when the learned decision rule relies on superficial correlation rather than the intended intrinsic features. For example, in image classification, the classifier may learn to use the background of an object, instead of the object itself for classification~\cite{bias}: A cow can be recognized if it appears in a typical cow-background like grass, but misclassified if put into an unexpected background like a beach. This behavior is analogous to our case where the two-tower model learns relevance from the position of the document rather than the (query, doc) features.\par 
Neural networks take shortcut because they find some unreliable features useful to cheaply optimize the target function. One intuitive way of discouraging such behavior is to make the shortcut features less useful for the target, e.g., by adding some random noise to the shortcut features.\par 
Following this intuition, we discourage the shortcut learning behavior by zero-out some neurons in the observation tower. Ideally, we want to only mask out relevance-related neurons for better interpretability and possibly better performance guarantee. However, we find a simple dropout layer~\cite{srivastava2014dropout} on the observation tower output works well empirically:
\begin{equation}
\label{eq:dropout}
  \psi(f, g) = {\rm sigmoid}(f + {\rm Dropout}(g, \tau)),
\end{equation}
where $\tau$ is the dropout probability.

\section{Theoretical Analysis} \label{sec:theory}
In this section, we theoretically analyze  how our proposed methods can reduce the confounding effect in the observation tower. 

The high-level intuition behind the theory is as follows: when the relevance features and observation features are correlated, the two-tower architecture has multiple optima minimizing the cross entropy with click labels, forming a continuous basin in the parameter space. However, only the global optimum when relevance and observation are uncorrelated corresponds to the optimal ranking performance with relevance labels. The vanilla two-tower model can only find an local optimum as a result of differences in learning efficiency of the two towers. Our methods, Gradient Reversal and Observation Dropout, both apply a tunable constraint on the learning efficiency of the two towers and can thus approach a subset of cross entropy optima, the best of whose ranking metric is better or as good as the optimum reached by the vanilla model. 

\subsection{Problem formulation}
Without loss of generality, we denote relevance related features as $\vec{x}^R$, and observation bias related features, such as positions and/or some relevance features~\cite{zhuang2021cross}, as $\vec{x}^O$, so that the click probability depends on the combined features as,
\begin{equation}
\label{eq:factorize}
P(C=1 | \vec{x}^R, \vec{x}^O) = P_{rel}(\vec{x}^R)\times P_{obs}(\vec{x}^O)
\end{equation}

Let's assume that we could perfectly model the click above with DNN models with the two-tower architecture with click probability,
\begin{equation}
\label{eq:model}
\hat{c} = f_\theta(\vec{x}^R) \times g_\varphi(\vec{x}^O),
\end{equation}
where $\theta$ and $\varphi$ are the parameters of the relevance tower and the observation bias tower, respectively.

As such, there is a set of parameters $\theta^*$ and $\varphi^*$ so that $f_{\theta^*}(\vec{x}^R) = p_{rel}(\vec{x}^R)$ and $g_{\varphi^*}(\vec{x}^O) = p_{obs}(\vec{x}^O)$ for any $\{\vec{x}^R, \vec{x}^O\}$. It is then straightforward to show that $(\theta^*, \varphi^*)$ optimize the average of the total loss on clicks in Eq.(\ref{eq:logloss}),
\begin{equation}
\label{eq:avgloss}
\langle\mathcal{L}^t\rangle = -\int dP^t(\vec{x}^R, \vec{x}^O)
\left[p(C=1)\log(\hat{c}) + p(C=0) \log(1-\hat{c})\right],
\end{equation}
where $dP^t(\vec{x}^R, \vec{x}^O)$ denotes the probability measure at $\{\vec{x}^R, \vec{x}^O\}$ of the logged data generated from the previous model $t$.

It's also easy to show that the corresponding model $f_{\theta^*}$ also optimizes the average relevance metrics like NDCG,
\begin{equation}
\label{eq:metrics}
\langle \mathcal{M}\rangle = \int dP(\vec{x}^R)m(p_{rel}(\vec{x}^R), f_\theta(\vec{x}^R)),
\end{equation}
where the metric function $m(y, \hat{y})$ is optimal when the model prediction is equal to the ground truth label, $\hat{y}=y$.
Similar to above, $dP(\vec{x}^R)$ denotes the marginal probability measure of the ranking candidates near $\vec{x}^R$.

\subsection{Confounding effect}
As we have discussed, in real-world data log, there exists a nontrivial correlation between positions and relevance features, in other words, $dP^t(\vec{x}^R, \vec{x}^O)\neq dP(\vec{x}^R)dP(\vec{x}^O)$. In the following, we will show and define how this correlation could negatively impact the two-tower model performance. 

We denote the marginal probability measure $dP(\vec{x}^R)$ is nonzero in the input space $\mathbb{R}$, and similarly, $dP(\vec{x}^O)$ is nonzero in the input space $\mathbb{O}$. In the following, we assume $\mathbb{R}$ and $\mathbb{O}$ are continuous differentiable domains for deep neural networks. For example, they could be representations of query-document pairs or numerical positions extended to the real numbers.

\begin{conjecture} \label{conj:1}
For the correlated $\vec{x}^R$ and $\vec{x}^O$, we then have the joint probability measure $dP^t(\vec{x}^R, \vec{x}^O)$ nonzero on the input space $\mathbb{X}^t$, which must be a subset of $\mathbb{R}\otimes\mathbb{O}$, i.e. $\mathbb{X}^t\subseteq\mathbb{R}\otimes\mathbb{O}$.
\end{conjecture}

\begin{theorem}\label{theorem:1}
If there is a parameter set different from the ground truth set $(\theta^*, \varphi^*)$ minimizing the loss function $\langle\mathcal{L}^t\rangle$ in Eq.(\ref{eq:avgloss}) on the input domain which is a strict subset of $\mathbb{R}\otimes\mathbb{O}$, then there exists a continuous set of parameter sets minimizing $\langle\mathcal{L}^t\rangle$: if $\exists(\theta, \varphi)\neq(\theta^*, \varphi^*)$ s.t. $\langle\mathcal{L}^t(\theta, \varphi)\rangle = \langle\mathcal{L}^t(\theta^*, \varphi^*)\rangle$, then $\exists\Theta=\{(\theta, \varphi) | \langle\mathcal{L}^t(\theta, \varphi)\rangle = \langle\mathcal{L}^t(\theta^*, \varphi^*)\rangle\}\supset\{(\theta^*, \varphi^*)\}$ s.t. $\forall\varepsilon>0$ and $\forall(\theta, \varphi)\in\Theta$, $\exists(\theta', \varphi')\neq(\theta, \varphi)$ s.t. $||(\theta', \varphi') - (\theta, \varphi)||<\varepsilon$ and $(\theta', \varphi')\in\Theta$.
\end{theorem}
\begin{proof} Assuming there is another parameter set, $(\theta_1, \varphi_1)\neq(\theta^*, \varphi^*)$, minimizing the average total loss in Eq.(\ref{eq:avgloss}), $\langle\mathcal{L}^t(\theta_1, \varphi_1)\rangle = \langle\mathcal{L}^t(\theta^*, \varphi^*)\rangle$. It's easy to see that $\langle\mathcal{L}^t(\theta_1, \varphi_1)\rangle = \langle\mathcal{L}^t(\theta^*, \varphi^*)\rangle$ if and only if $f_{\theta_1}(\vec{x}^R)g_{\varphi_1}(\vec{x}^O) = p_{rel}(\vec{x}^R)p_{obs}(\vec{x}^O)$ for $\forall\{\vec{x}^R, \vec{x}^O\}\in\mathbb{X}^t$. Let's say $f_{\theta_1}(\vec{x}^R)g_{\varphi_1}(\vec{x}^O) = p_{rel}(\vec{x}^R)p_{obs}(\vec{x}^O)$ on $\mathbb{X}_1$ so that $\mathbb{X}^t\subseteq\mathbb{X}_1\subset\mathbb{R}\otimes\mathbb{O}$. Due to the continuity of the Deep neural network on parameters $(\theta, \varphi)$, there must exist a neighbor $(\theta_2, \varphi_2)$ in any close vicinity of $(\theta_1, \varphi_1)$ so that $f_{\theta_2}(\vec{x}^R)g_{\varphi_2}(\vec{x}^O) = p_{rel}(\vec{x}^R)p_{obs}(\vec{x}^O)$ on $\mathbb{X}_2$ with $\mathbb{X}_1\subseteq\mathbb{X}_2\subseteq\mathbb{R}\otimes\mathbb{O}$, where the second equal sign is true if and only if $(\theta_2, \varphi_2)=(\theta^*, \varphi^*)$.
\end{proof}



\begin{proposition}~\label{prop:1}
 The relevance metrics are suboptimal for the parameters in the loss optimal set different from the ground truth $(\theta^*, \varphi^*)$.
\end{proposition}
\begin{proof} $\langle \mathcal{M}\rangle$ is maximal if and only if $f_\theta(\vec{x}^R) = p_{rel}(\vec{x}^R) = f_{\theta^*}(\vec{x}^R)$ for $\forall\vec{x}^R\in\mathbb{R}$, as $\frac{\delta^2}{\delta f_\theta(\vec{x}^R)^2} m(p_{rel}(\vec{x}^R), f_{\theta}(\vec{x}^R))<0$ at $f_{\theta^*}(\vec{x}^R)$. For a non-constant function $f_\theta(\vec{x}^R)$, its value does not equal to $f_{\theta^*}(\vec{x}^R)$ for $\forall\vec{x}^R\in\mathbb{R}$ if $\theta\neq\theta^*$. So for any $(\theta,\varphi)\in \Theta\textbackslash\{(\theta^*, \varphi^*)\}$, $\langle\mathcal{M}(\theta)\rangle<\langle\mathcal{M}(\theta^*)\rangle$.
\end{proof}

\begin{postulate}\label{post:1}
Click dependence on the confounding features are easier to be fitted with neural network model in the observation tower $g_\varphi$ than in the relevance tower $f_\theta$, so that when optimizing the two-tower model Eq.(\ref{eq:model}), the model converges to a parameter set $(\theta^{\rm sub}, \varphi^{\rm sub})\neq(\theta^*,\varphi^*)$ with $\langle\mathcal{L}(\theta^{\rm sub}, \varphi^{\rm sub})\rangle = \langle\mathcal{L}(\theta^*, \varphi^*)\rangle$.
\end{postulate}

\begin{corollary}\label{coro:1}
The relevance metric $\langle \mathcal{M}\rangle$ is suboptimal for a two-tower model trained on the clicks generated from correlated relevance and observation bias satisfying the above Postulate 4: $\langle\mathcal{M}(\theta^{\rm sub})\rangle < \langle\mathcal{M}(\theta^*)\rangle$.
\end{corollary}
\begin{proof} 
Applying Theorem~\ref{theorem:1} and Proposition~\ref{prop:1} to Postulate~\ref{post:1}, we can derive the above Corollary.
\end{proof}



\subsection{Application to Our Methods}
Given that the correlation in the relevance inputs $\vec{x}^R$ and the observation inputs $\vec{x}^O$ degrades the two-tower model performance on the relevance metrics, we will discuss how each of our method could help recover the relevance predictions of the two-tower models.

\begin{proposition}[Dropout method]
\label{prop:dropout}
Given a dropout rate $\tau$, the model converges to an optimum $(\theta_\tau,\varphi_\tau)\in\Theta$ different from $(\theta^{\rm sub}, \varphi^{\rm sub})$ when $\tau>0$. The relevance metrics $\langle\mathcal{M}(\theta_\tau)\rangle$ can be optimized at $\tau^*\in[0, 1]$ {\it s.t.} $\langle\mathcal{M}(\theta_{\tau^*})\rangle\geq\langle\mathcal{M}(\theta^{\rm sub})\rangle$.
\end{proposition}

To prove the above Proposition \ref{prop:dropout}, we first need to show,
\begin{lemma}
\label{lemma:dropout1}
Parameters $(\theta, \varphi)$ optimizing the average loss function with a dropout $\tau>0$ also optimize the original loss function $\langle\mathcal{L}\rangle$ in Eq.(\ref{eq:avgloss}): $\Theta^{\rm dropout} = \{(\theta_\tau, \varphi_\tau)|0\leq\tau<1\}\subseteq\Theta$.
\end{lemma}
\begin{proof}
By variation at $\{\vec{x}^R, \vec{x}^O\}$, we can see that the average loss is optimized only when $p(C=1|\vec{x}^R, \vec{x}^O) = \overline{\hat{c}}^{\tau} = f_{\theta_\tau}(\vec{x}^R)g_{\varphi_\tau}(\vec{x}^O)$ for $\forall\{\vec{x}^R, \vec{x}^O\}\in\mathbb{X}^t$. So $(\theta_\tau, \varphi_\tau)\in\Theta$.
\end{proof}

\begin{lemma}
\label{lemma:dropout2}
For the set of parameters optimizing the dropout method, $\Theta^{\rm dropout}$, the optimal relevance performance must be better or at least as good as the performance of the vanilla two-tower model: $\sup_{\Theta^{\rm dropout}}\langle\mathcal{M}(\theta_\tau)\rangle\geq\langle\mathcal{M}(\theta^{\rm sub})\rangle$.
\end{lemma}
\begin{proof}
It's straightforward to show that there exists $\theta \in \Theta^\theta$ with $\langle\mathcal{M}(\theta)\rangle > \langle\mathcal{M}(\theta^{\rm sub})\rangle$, as $\theta^*\in\Theta^\theta$ and $\langle\mathcal{M}(\theta^*)\rangle > \langle\mathcal{M}(\theta^{\rm sub})\rangle$ in Corollary~\ref{coro:1}. So if there exists such a $\theta\in\Theta^{\rm dropout,\theta}$, the greater sign would be satisfied. On the other hand, $\theta_{\tau=0}=\theta^{\rm sub}\in\Theta^{\rm dropout,\theta}$ by definition, so the equal sign would be satisfied if no such a $\theta$ with better relevance metric belongs to $\Theta^{\rm dropout}$.
\end{proof}

Based on Lemma~\ref{lemma:dropout1} and Lemma~\ref{lemma:dropout2}, we can derive Proposition~\ref{prop:dropout}, which guarantees that we can find a two-tower model with a relevance performance as good or better. In practice, by dropping out the observation prediction, to which the model overly attributes the dependence of confounding features, we move the converging optimum in the direction towards the ground truth optimum. The trained model is thus able to approach to a parameter set with relevance ranking performance closer to the ground truth ${\theta^*}$.

\begin{proposition}[Gradient reversal method]
\label{prop:gradrev}
Given a gradient reverse coefficient $\eta$, the model converges to an optimum $(\theta_\eta,\varphi_\eta)\in\Theta$ different from $(\theta^{\rm sub}, \varphi^{\rm sub})$ when $\eta>0$. The relevance metrics $\langle\mathcal{M}(\theta_\eta)\rangle$ can be optimized at $\eta^*\in[0, \infty)$ {\it s.t.} $\langle\mathcal{M}(\theta_{\eta^*})\rangle\geq\langle\mathcal{M}(\theta^{\rm sub})\rangle$.
\end{proposition}

Proposition~\ref{prop:gradrev} can be derived in a similar way as Proposition~\ref{prop:dropout}. Similar to Proposition~\ref{prop:dropout}, Proposition~\ref{prop:gradrev} guarantees the relevance performance of the gradient reversal method to be equal or better than that of the vanilla two-tower model. In the situation that the vanilla model over-depends on the confounding features in the observation tower. Reversed gradients hinder the learning of such dependence, and thus lead to a relevance performance closer to the ground truth $f_{\theta^*}$ at $\eta^*>0$.

Despite that all the above proofs rely on the strong assumptions of the perfect solvability of the DNN models to the relevance and observation functions and continuity of the input space, relaxing these assumptions still allows most of the conclusions to be valid in practice as we will test in experiments.

\section{Experiments}
In this section, we discuss our experiment setting and validate our methods offline on two public learning to rank datasets with synthetic clicks and a large-scale industrial dataset with logged user interactions from the Google Chrome Web Store (CWS). We also perform online A/B experiments of the proposed methods on the CWS production system.

\subsection{Methods}
We compare the following methods on two semisynthetic datasets and the real-world CWS dataset.
\begin{itemize}
    \item Biased baseline (\textbf{Biased}): A single tower feed-forward neural network model that takes only the regular (query, doc) features trained to predict the biased clicks. No observation-bias related features are included.
    \item Two Tower Additive Model (\textbf{PAL})~\cite{guo2019pal}: The standard unbiased model that consists of two towers of feed-forward neural networks: a relevance tower takes regular (query, doc) features to model relevance predictions and the other observation tower takes observation-bias related features like positions. The outputs of both towers are added together to predict user click probability logit.
    \item Dual Learning Algorithm (\textbf{DLA}) \cite{ai2018unbiased}: DLA jointly trains a ranking model and an examination propensity estimation model with an assumption of duality between relevance learning and propensity examination. In our experiment, the ranking model is a DNN same as our \textbf{Biased} baseline.
    \item Gradient Reversal (\textbf{GradRev}): The two-tower baseline with a gradient reversal layer applied to the observation tower. See method details in Sect.~\ref{sec:gradrev}. In the results below, we always present the results from using clicks as the adversarial labels. Results for other adversarial labels are discussed in the Ablation study in Sect.~\ref{sec:ablation-study}. 
    \item Observation Dropout (\textbf{Drop}): The two-tower baseline with a dropout layer applied to the output of the observation tower. See method details in Sect.~\ref{sec:dropout}.
\end{itemize}

\subsection{Semisynthetic Datasets Setup}
\paragraph{LTR Datasets}
We use Yahoo Learning to Rank Set1 (Yahoo)~\cite{yahoo} and MSLR-WEB30K Fold1 (Web30k)~\cite{web30k} to benchmark our methods. Yahoo contains 19944 training queries, 2994 validation queries, and 6983 test queries. Each query is associated with 20 documents on average. Each query-doc contains 700 dense features. Web30k contains 31531 queries, divided into train, validation, test set with a ratio of 3:1:1. Each query on average has 120 documents, with 136 dense features for each query-doc pair. Both datasets are labeled with 5-grade relevance judgement from 0 to 4.\par

\paragraph{Logging Policy}
Our goal is to understand how the confounding between relevance and observance modeling would affect the relevance learning, and how we can mitigate such effects. Toward this goal, we design  logging policies to generate observation features of different quality in terms of their correlation with the relevance, and test their effects on the relevance learning in two-tower models by measuring the relevance tower performance. 

Intuitively, observation is mostly entangled with relevance when positions are fully determined by ground-truth relevance scores. So we consider an \emph{oracle} logging by ranking all the documents of each query based on their relevance scores. We expect such setting would guarantee the worst relevance tower performance in the two-tower model.
At the other end, the observation bias is least entangled with relevance when input positions are randomly shuffled despite the query-doc features. We expect such setting would set the upper bound for the relevance tower performance. 
We also interpolate the cases between the extremes using a mixture of relevance-based ranking and random shuffling.

In particular, we examine 5 different logging policies in our experiments using the methods described above. For each document $d_i$, we have a ground-truth relevance label, $y_i$, and a random noise, $n_i$, extracted from a uniform distribution in [0,4]. We assign a weight $w$ to $y_i$, and $(1-w)$ to ${n_i}$. We rank all the documents of a query based on the descending order of a weighted sum score $s_i = wy_i + (1-w)n_i$ and use the rank as the logged position $p_i$ for $d_i$.
\begin{itemize}
\item Oracle: $w=1.0$ 
\item L1: $w=0.8$ 
\item L2: $w=0.6$ 
\item L3: $w=0.2$ 
\item Random: $w=0.0$ 
\end{itemize}
\vspace{-0.6em}
\paragraph{Generating Synthetic Clicks}
Given the logged positions, we use a common position bias click model (PBM) to simulate user clicks, following previous studies~\cite{richardson2007predicting}. Click probability is given by the product of relevance score and observation probability. Observation probability is assumed to be only related to its position $p_i$ as:
\begin{equation}
\label{eq:observation}
    P(O_i=1|p_i) = \frac{1}{p_i}
\end{equation}
Once a document is examined, user will click based on its relevance label, $y_i$, but with certain noise, $\epsilon$, in the decision. We consider the click probability that users find a document relevant as:
\begin{equation}
\label{eq:relevance}
    P(R_i=1|y_i) = \epsilon + (1-\epsilon)\frac{2^{y_j}-1}{2^{y_{max}}-1}
\end{equation}
where $y_{max}=4$ is the maximum relevance score in the datasets and $\epsilon$ is the noise constant, which we set to 0.1. The click probability of a document of relevance $y_i$ appearing at position $p_i$ is:
\begin{equation}
P(c_i=1|y_i,p_i) = P(O_i=1|p_i)\times P(R_i=1|y_i)
\end{equation}

\subsection{Results of Semisynthetic Datasets}
We evaluate the relevance performance of two-tower models by checking the NDCG@5 of the relevance tower predictions on the ground-truth relevance label $y_i$. Main results are shown in  Table \ref{tbl:ndcg-syn}.\par


\begin{table}[ht]
\centering
\caption{Relevance prediction performance on Yahoo LTR and Web30k, measured by NDCG@5 of utility label. Best results are bolded. Significant improvement ($\alpha=0.05$) of the method over the PAL baseline is marked by upperscript *. }
\vspace{-1em}
\label{tbl:ndcg-syn}
\resizebox{0.48\textwidth}{!}{
\begin{tabular}{@{} p{1cm} *{6}{>{\rmfamily}l} @{}}
\toprule[.1em]
Dataset & Logging  & Biased &  PAL & DLA &  GradRev  &  Drop \\
\midrule[.1em]
  \multirow{5}{*}{Yahoo} & Oracle & 0.7048 & 0.6836  &  0.6754 &  $ 0.7126^*$ & $\mathbf{0.7157^*}$\\
  \addlinespace
  & L1 & 0.7038 & 0.6831 & 0.6893 & $0.7149^*$ & $\mathbf{0.7159^*}$\\
  \addlinespace
  & L2 & 0.6929  & 0.7058 & 0.7034 & $0.7162^*$ & $\mathbf{0.7169^*}$\\    
  \addlinespace
  & L3 & 0.6837& 0.7140  & 0.7030 & $\mathbf{0.7147^*}$ & 0.7145\\ 
  \addlinespace
  & Random & 0.6630 & 0.7179 & 0.7094 & 0.7189 & $\mathbf{0.7199^*}$\\  
\midrule[.1em] 
  \multirow{5}{*}{Web30k} 
  & Oracle & 	$\mathbf{0.4191}$ & 0.3333 & 0.3144 & $0.4159^* $ & $0.4118^*$  \\ 
  \addlinespace
  & L1 & 0.3984 & 0.2939 & 0.3222 & $\mathbf{0.4221^*}$ & $0.4087^*$  \\ 
  \addlinespace
  & L2 & 	0.3878 & 0.3166 & 0.3435 &$\mathbf{0.4334^*}$ & 	$0.4130^*$ \\     
  \addlinespace
  & L3 & 0.3880 &0.3524& 0.3612 & $0.4119^*$ & $\mathbf{0.4222^*}$ \\    
  \addlinespace
  & Random & 0.3860 & 0.4103  & 0.3755 &$\mathbf{0.4140}$ & 0.4056  \\ 
\bottomrule[.1em]
\end{tabular}
}
\end{table}

First, compared to the Random case with independent relevance and position, the baseline two-tower PAL model performance degrades as the logging policy involves more and more relevance information from L3 to Oracle. This result aligns with our hypothesis that the confounding between relevance and observation learning has negative effects on the relevance tower.\par
On the contrary, the biased model with no position input show improving performance, as the more correlation between relevance and position leads to more clicks and thus more information in the synthetic clicks. But in all cases, the biased model performance is always significantly worse than the two-tower model upper bound obtained with the independent relevance and position.\par
We find that both of the proposed methods have improvements over all the baselines. Especially, we find larger improvements in the proposed models when the observation feature is more correlated with the relevance: The improvement goes up to 4.7\% in Yahoo as the logging policy is perfectly correlated with the relevance. The best performance of each method stays close to the two-tower model upper bound, rather independent of the logging policy.\par
Comparing the disentangling methods, we notice our proposed two methods consistently give at least on par, if not better result across different logging policies.\par 

\subsection{Results of an Industrial Dataset}
\begin{figure}[t]
\includegraphics[scale=0.13]{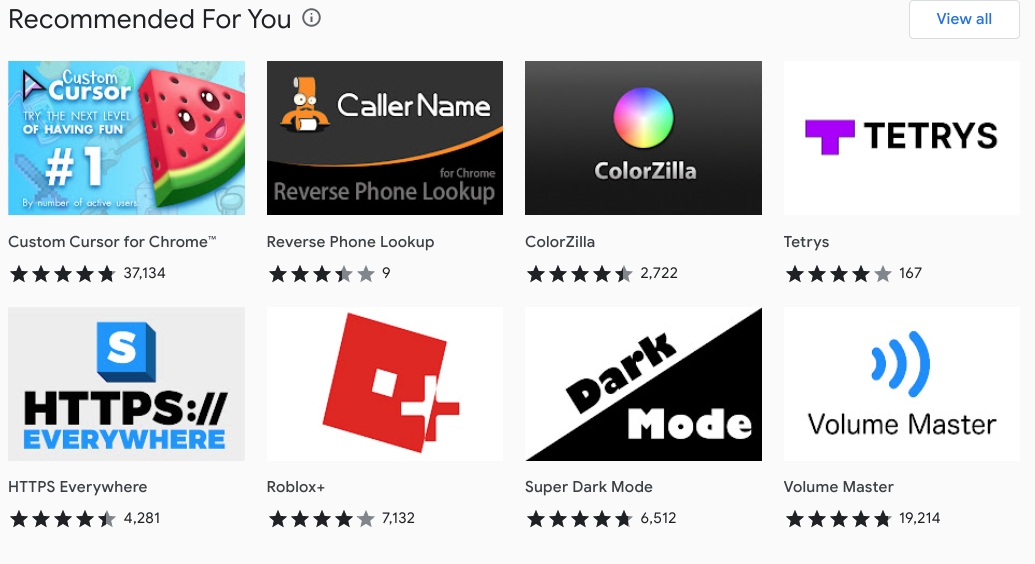}
\vspace{-0.8em}
\caption{An example of Chrome Web Store user interface.}
\label{fig:cws-ui}
\end{figure}
\paragraph{Dataset}
\begin{table}[bh!]
	\caption{Ranking performance on the CWS dataset, measured by NDCG@5, and IPS-NDCG@5 of installs and clicks. Best results for each metric are bolded. Significant improvement ($\alpha=0.05$) over the PAL baseline is marked by *.}
	\centering
	\resizebox{0.45\textwidth}{!}{
	\begin{tabular}{c|cccc cccc}
     \hline
	\multirow{2}{*}{\textbf{Model}}&\multicolumn{2}{c}{\textbf{Install}}&\multicolumn{2}{c}{\textbf{Click}}\\
	&\multicolumn{1}{c}{\textbf{NDCG}}&\multicolumn{1}{c}{\textbf{IPS-NDCG}}&\multicolumn{1}{c}{\textbf{NDCG}}&\multicolumn{1}{c}{\textbf{IPS-NDCG}}\\
			\hline
			Biased & 0.3135 & 0.3077  & 0.5045  & 0.4884\\
			PAL & 0.3108 & 0.3066 & 0.4945& 0.4825\\
			GradRev &0.3144 &$\mathbf{0.3104^* }$&0.5050 & 0.4890 \\
			Drop &$\mathbf{0.3150^*}$ &0.3103 &$\mathbf{0.5056^*}$ & $\mathbf{0.4901^*}$\\
     \hline
	\end{tabular}}
\label{tbl:ndcg-cws}
\end{table}
We also test the methods on the Chrome Web Store (CWS) dataset~\cite{bootstrapping2021qin}, which contains logged display history of Chrome Web Store Homepage (see \figref{fig:cws-ui} for the user interface) and user behaviors, including both click and install. The dataset contains several relevance-related features, e.g., average recent rating, and several observation-related features, e.g., layout index. 
The logs are collected for one month over the October (27 days) and November (3 days) in 2021. We train models with the October's logs only on logged clicks and test the performance on the November's logs. After trained offline, we  deploy the trained models in an online A/B testing on Chrome Web Store homepage recommendation for four weeks in January 2023.\par 

\paragraph{Offline Results}
 We summarized the results in Table \ref{tbl:ndcg-cws}. As we do not have ground-truth relevance labeled for CWS, we evaluate the relevance tower predictions on both logged clicks and installs using both NDCG  metric and NDCG metric corrected with the Inverse Propensity Scores, (IPS NDCG), where the IPS is calculated with the average clicks at a give position. We observe that all the methods improves the relevance tower performance over the vanilla two-tower PAL baseline, bringing around 1.3\% improvement, which is considered significant in the production setting. This result indicates that the confounding effects could be general in real-world user logs and our proposed methods could further boost the real-world two-tower models in such situations.\par 

\paragraph{Online A/B testing} Due to resource constraints, we only test three experiments methods: PAL, Biased, and GradRev. PAL is the standard two-tower debiased model used in production. We report relative percentage change over PAL in \tblref{tbl:online}.

\begin{table}[h]
\caption{Performance comparison in online A/B expreriments on Chrome Web Store, compared against the production PAL model.}
\vspace{-0.8em}
\begin{tabular}{c|c|c} 
  \hline
   Metrics & Click rate & Users with clicks  \\ 
  \hline
  Biased & $+1.21\%$ & $-5.78\%$ \\ 
  GradRev   & $+2.90\%$ & $+0.96\%$ \\
  \hline
\end{tabular}
\label{tbl:online}
\end{table}

Compared to the PAL baseline, the Biased single tower model gains some clicks, but with fewer users to click. The result indicates a strong bias towards extensions that have already gained a lot of clicks in some users, as a result, the biased recommendations gain more clicks in these users at the same time lose the clicks in the rest.
Instead, the debiased GradRev model using gradient reversal to reduce the confounding effects gains in both total clicks and total user with clicks, which implies less bias and higher quality in the recommended extensions. The online user performance improvement of GradRev is considered significant in the application.

\subsection{Ablation Study}
\label{sec:ablation-study}

\paragraph{Adversarial Label Study}
For the gradient reversal method, we study how the choice of adversarial labels affects the method efficacy. 
Given the fact that production logs usually don't contain ground-truth relevance labels, we experiment several alternatives, including clicks (Click), predicted relevance score (Prediction) by a pre-trained relevance tower. We compare them in training with ground-truth utility (Utility) as the adversarial label, which is assumed to set the upper-bound for the gradient reversal method. 
\vspace{-0.2em}
\begin{table}[h]
\caption{Performance comparison for the gradient reversal method on Yahoo, given different adversarial labels.}
\vspace{-1em}
\begin{tabular}{ c c c c } 
  \hline
  
 
  
  
  Logging  & Utility & Click & Prediction  \\ 
  \hline
  Oracle & 0.7150 & 0.7126 
 & \textbf{0.7165}\\ 
  
  L1 & 0.7128 & 	\textbf{0.7149}  & 0.7092  \\ 
 
  L2 & \textbf{0.7186} & 0.7162 & 0.7149\\ 
  
  L3 &  0.7150 & 0.7147 
 & \textbf{0.7151} \\ 
  
  Random & 0.7179  & 0.7189 & \textbf{0.7200} \\ 
  \hline
\end{tabular}
\label{tbl:comparison-gradrev-yahoo}
\vspace{-0.5em}
\end{table}

From Table~\ref{tbl:comparison-gradrev-yahoo}, we find that the three adversarial labels achieve comparable performance without any of them getting an obvious edge. This observation indicates that the method is rather robust to the choice of the adversarial labels. This promises the applicability of the method in wide scenarios when the ground-truth relevance labels are not available.

\begin{figure}
\centering
\includegraphics[width=7cm]{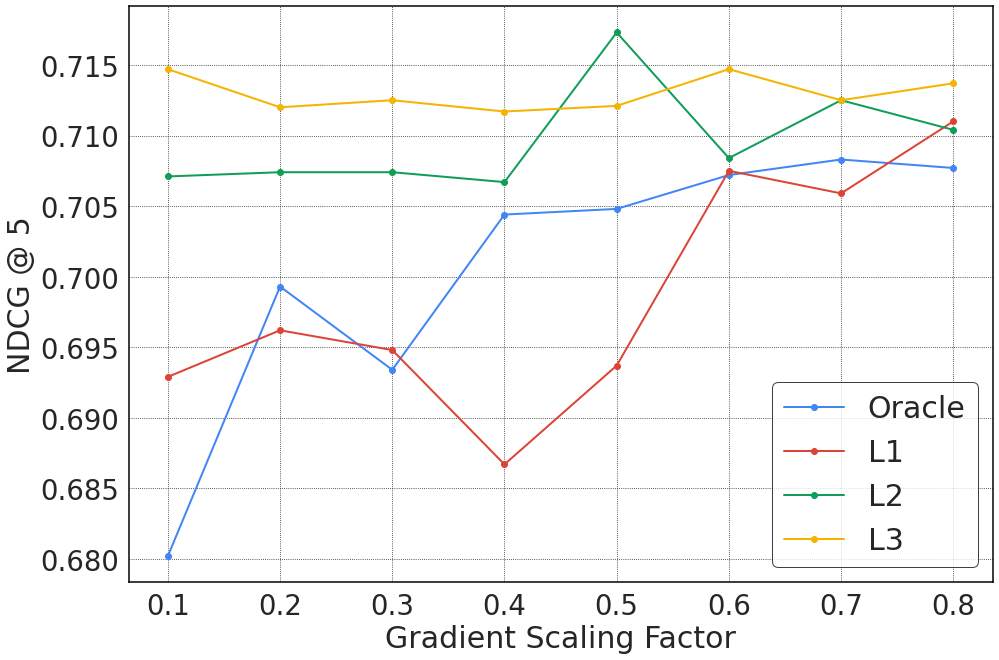}
\vspace{-1em}
\caption{Performance comparison between different gradient scaling rates for the gradient reversal method.}
\label{fig:perf-scaling}
\end{figure}

\begin{figure}
\centering
\includegraphics[width=7cm]{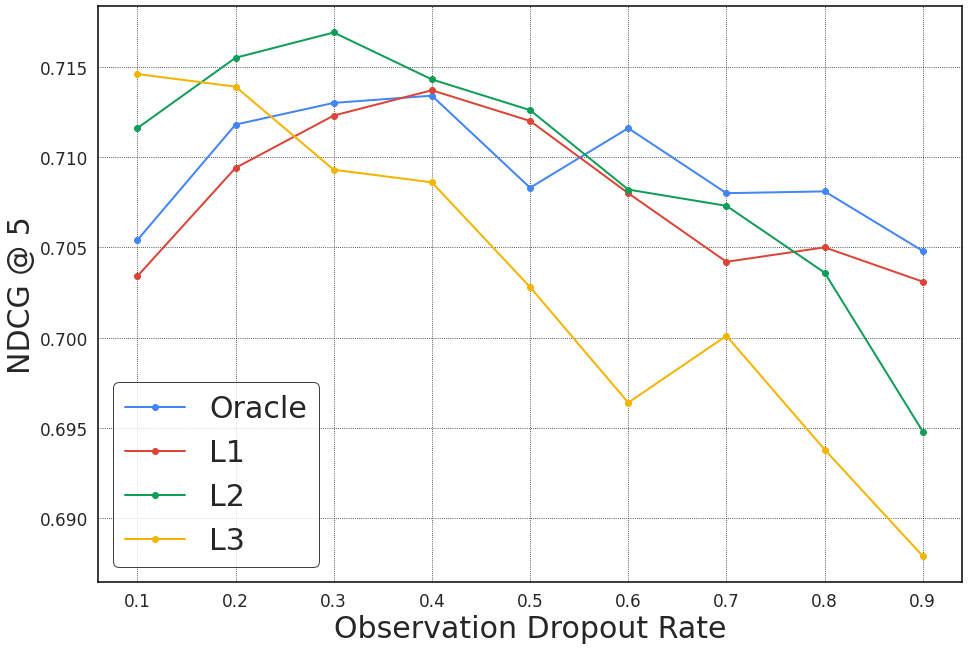}
\vspace{-1em}
\caption{Performance comparison between different observation dropout rates for the observation dropout method.}
\label{fig:perf-dropout}
\end{figure}

\paragraph{Sensitivity Study}
We analyze the sensitivity of the Gradient Reversal method and the Observation Dropout method against their hyper-parameters, as illustrated in Figure~\ref{fig:perf-scaling} and Figure~\ref{fig:perf-dropout} for Yahoo.  
For the Gradient Reversal method, shown in Figure~\ref{fig:perf-scaling}, the performance generally increases with a larger gradient scaling factor and takes optimal around 0.6 to 0.8, when the input position becomes strongly correlated with the relevance in Oracle and L1. At the same time, the performance is rather less sensitive to the choice of the gradient scaling factor in the range we test when the correlation becomes weak, as in L3. Observation Dropout performance, shown in Figure~\ref{fig:perf-dropout}, presents a non-monotonic dependence on the dropout rate: the model takes optimal performance between a dropout rate of 0.2 and 0.5. The other trend we observe is that the more correlation between the position and the relevance in the logging policy, we need to search for a larger dropout rate $\alpha$ on the observation tower prediction to optimally disentangle them.

\paragraph{Method Combination}
Finally, we study the effects of combining our proposed tricks. 
We investigate the combination of the proposed methods in the oracle setting, compared to the best score we achieved using a single method. We can see Drop + GradRev can get on par or even slightly improve the performance by $0.5\%$. This result aligns with our theoretical analysis: the Dropout method and the Gradient Reversal method can be unified in the same framework and thus may work additively.

\section{Related Work}
The two-tower model family for ULTR has been widely explored in both industry and academia. PAL~\cite{guo2019pal} is the pivot work that introduces the two-tower model to the research community. \citeauthor{haldar2020improving}~\cite{haldar2020improving} apply a two-tower model on Airbnb search. \citeauthor{feeds}~\cite{feeds} adapts a two tower model for news feed recommendation ranking with a rich set of bias features. \citeauthor{zhuang2021cross}~\cite{zhuang2021cross} considers surrounding items in the observation model.Similarly, \citeauthor{gnnrec}~\cite{gnnrec} models the neighboring items with GNN. \citeauthor{mhrec}~\cite{mhrec} further enhances the two-tower model with self-attention. \citeauthor{revisit}~\cite{revisit} enriches two-tower models with more expressive interactions and click models, and is orthogonal to our work. \citeauthor{youtube}~\cite{youtube} uses the two-tower model for recommending which video to watch on YouTube. Their model uses a dropout operation that look similar to one of our proposed methods. However, their dropout is added to the \textit{input} layer to learn a default position representation and do not consider the confounding effect. \citeauthor{lbd}~\cite{lbd} studies the dropout method to adjust the debiasing strength according to individual-level coupling of feature-observation pairs. Our work focus on resolving the entanglement between relevance and observation bias  itself. To the best of our knowledge, no prior work discusses the general confounding effect caused by the ranking process itself.

Another closely related family of methods for ULTR, Inverse Propensity Scoring (IPS) based methods~\cite{joachims2017unbiased, wang2018position, ai2018unbiased, agarwaL2019general, wang2016selectionbias, oosterhuis2020policy, attribute, wsdm2021,vector,mou} follow the same Position Based Model assumption (except for very few recent works~\cite{cmunbiased}). In this work, we focus on discussing two-tower models due to their popularity, but the discussed concerns may generalize to IPS-based methods since they follow the same assumptions. For example, \citeauthor{wang2018position}~\cite{wang2018position} performs an Expectation-Maximization procedure between relevance sub-model and observation propensity sub-model. Though the goal was to estimate propensity, the confounding between these two sub-models persists and the propensity estimation can be negatively affected. Studying the confounding effect in these methods is left for future work.

\section{Discussion}
Here we discuss several limitations in our work for the follow-up study. Firstly, our study relies on the assumption that each user operates independently for the sake of simplicity. However, this is not always the case in sectors such as e-commerce, where users are influenced by the feedback of their peers, such as ratings and comments. Consequently, user behavior is not strictly independent, but somewhat interdependent. It would be beneficial to examine how the relaxation of the user-independence assumption might refine our theoretical analysis.

Secondly, there is a tradeoff to consider between the performance gain and training cost for the gradient reversal method. Our experimental results indicate that the gradient reversal method tends to converge at a slightly slower pace, with a more volatile training curve compared to the baselines. Thus, we suggest further optimizations to improve the training efficiency of this method.
\section{Conclusion}
In this work, we re-examine the factorization assumption in unbiased learning to rank frameworks. We show that the confounding between relevance modeling and observation bias modeling can hurt relevance predictions in controlled semisynthetic datasets. We also propose two effective methods, gradient reversal and observation dropout to alleviate the negative effects. We demonstrate that the proposed methods can achieve superior performance in two semisynthetic datasets and a real-world dataset. Lastly, we theoretically show why the confounding issue could hurt model performance and how our methods work.

\bibliographystyle{ACM-Reference-Format}
\balance
\bibliography{references}

\end{document}